\documentclass{article}
\usepackage{amsfonts}
\usepackage{amsmath}
\usepackage{hyperref}
\usepackage{curves}
\usepackage{enumerate}
\usepackage[utf8]{inputenc}
\usepackage{latexsym}
\usepackage{braket}
\usepackage{mathrsfs}
\usepackage{mathtools}
\usepackage{color}
\usepackage{xcolor}
\usepackage{enumitem}
\usepackage[a4paper, total={6in, 8in}]{geometry}

\usepackage{cite}
\usepackage[english]{babel}
\usepackage{babelbib}
\usepackage{multirow}
\usepackage{tabularx}
\usepackage{makecell}
\setcellgapes{4pt}
\usepackage[thinlines]{easytable}

\usepackage{colortbl}

\begin{document}

\title{\textbf{The solution space to the Einstein's vacuum field equations for the case of five-dimensional Bianchi Type I (Type $\mathbf{4A_{1}}$) }}
\author{ \textbf{T. Pailas}$^b$\thanks{teopailas879@hotmail.com} ,\,\,\textbf{Petros A. Terzis}$^b$\thanks{pterzis@phys.uoa.gr},\,  \textbf{T. Christodoulakis}$^b$\thanks{tchris@phys.uoa.gr}\\
\normalsize
{$^b$\it Nuclear and Particle Physics Section, Physics
Department,}\\
\normalsize{\it National and Kapodistrian University of Athens, GR 157--71 Athens, Greece}}

\maketitle

\abstract{We consider the 4+1 Einstein's field equations (EFE's) in vacuum, simplified by the assumption that there is a four-dimensional sub-manifold on which an isometry group of dimension four acts simply transitive. In particular we consider the Abelian group Type $4A_{1}$; and thus the emerging homogeneous sub-space is flat. Through the use of coordinate transformations that preserve the sub-manifold's manifest homogeneity, a coordinate system is chosen in which the shift vector is zero. The resulting equations remain form invariant under the action of the constant Automorphisms group. This group is used in order to simplify the equations and obtain their complete solution space which consists of seven families of solutions. Apart form the Kasner type all the other solutions found are, to the best of our knowledge, new. Some of them correspond to cosmological solutions, others seem to depend on some spatial coordinate and there are also pp-wave solutions.}

\newpage

\section{Introduction}

It is commonly known that the concept of symmetry possesses a fundamental role in mathematical physics. Specifically, in the branch of general relativity, symmetry has been used in order to simplify and subsequently solve (EFE's), as were as classify the solutions, see for example \cite{stephani_kramer_maccallum_hoenselaers_herlt_2003},\cite{Griffiths:2009dfa}. An interesting example is that of the group of Automorphisms also called rigid symmetries \cite{Coussaert:1993ti}. Ashtekar and Samuel were the first to study the group of Automorphisms from a geometric viewpoint \cite{0264-9381-8-12-005}. In the case of 3+1 Bianchi Types the use of Automorphisms of three dimensional Lie algebras has proven very useful since eventually leads to the specification of the general solution space for Bianchi Types (I-VII)\cite{Christodoulakis:2004yx},\cite{Christodoulakis:2006vi},\cite{Terzis:2008ev},\cite{Terzis:2010dk}. Furthermore, the Automorphisms provide an algorithm for counting the number of essential constants and therefore invariantly characterize the manifold at hand\cite{ellis1969}, \cite{0305-4470-36-2-310},\cite{Christodoulakis:2003wy}.

The existence of extra dimensions seems to appear in mathematical physics since the seminal works of Kaluza and Klein\cite{Kaluza:1921tu},\cite{Klein1926}. Their theory was a prototype for more sophisticated theories developed in later years, such as, string theory, brane theory, supergravity, supersymmetry e.t.c. We cite only a few articles dealing with these subjects since there are two many to include them all \cite{PhysRev.52.1107},\cite{Heisenberg1943},\cite{Nambu:1969se},\cite{Susskind:1970qz},\cite{SCHERK1974118},\cite{GREEN1984117},\cite{DUFF1988515},\cite{Randall:1999ee}. 

Much work has been done in the context of higher dimensional cosmology. In a paper of P. Forgacs and Z. Horvath \cite{Forgacs:1978xy}, homogeneous and isotropic universes in the presence of gauge fields and two extra compact dimensions were studied. The idea that the properties of matter in the four-dimensional universe can be purely described geometrically by using an extra dimension, works remarkably well in the case of spatially flat cosmological solutions in the presence of a perfect fluid, as was shown by Paul S. Wesson \cite{Wesson:1992hh}. The extension of this work in general spatially FRLW cosmological solutions was presented in \cite{1994JMP....35.4889M}. Also, a class of wave-like solutions were derived in \cite{Liu:1994if}. More recently, non-separable five-dimensional solutions in which the induced four-dimensional metric has the form of FRLW cosmology, were obtained in \cite{Fukui:2001en}. Alan Chodos and Steven Detweiler \cite{PhysRevD.21.2167} considered a five-dimensional extension of Kasner's four-dimensional solution \cite{10.2307/2370192},\cite{10.2307/1989167}. This was the first attempt to study anisotropic cosmological spaces in the presence of extra dimensions. A series of papers concerning higher dimensional anisotropic spacetimes in Einstein's general relativity, modifications of it like Brans-Dicke theory and supergravity, were presented by D.Lorenz-Petzold \cite{LorenzPetzold:1985wd}, \cite{LorenzPetzold:1985nk}, \cite{LorenzPetzold:1985jm}, \cite{LORENZPETZOLD1985105}, \cite{LorenzPetzold:1985vv}, \cite{Lorenz-Petzold1985}, \cite{LorenzPetzold:1985jr}, \cite{LorenzPetzold:1986bs}. The existence of chaotic solutions in some of the five-dimensional homogeneous spacetimes was studied by Paul Halpern \cite{Halpern:1986ue}. In the paper \cite{Hervik:2002af}, Sigbjørn Hervik classifies the five-dimensional cosmological models based on whether the spatial hypersurfaces are connected or simply connected homogeneous Reimannian manifolds. Finally, some homogeneous vacuum plane wave solutions of five-dimensional (EFE's) are obtained in \cite{0264-9381-20-19-312}.

In this work, we are interested in the case of five-dimensional manifolds possessing a four-dimensional homogeneous sub-manifold. Specifically, we will concentrate in the Type 4A1. This enumeration can be found in the works of \cite{doi:10.1063/1.523441},\cite{Terzis:2013ksa}. This Type is analogous to the four-dimensional Bianchi Type I, since all structure constants vanish and therefore the sub-manifold is flat. Thus, the group of Automorphisms is represented by the $GL(4,R)$ matrices. The corresponding Automorphisms for all the other real four-dimensional Lie algebras can be found in\cite{0305-4470-36-2-310}. Eventually, the Automorphisms will provide us a way to separate the different families of solutions to (EFE's) and find their solution space.  

The paper's structure is organized as follows. In section (II) an introduction to the canonical formalism and the homogeneous manifolds is provided alongside with the 4+1 form of Einstein's equations. The basic idea of transformations which preserve the manifest homogeneity is reviewed in section (III). In section (IV), everything mentioned in the previous sections is applied to the case of Type 4A1 and the way that the constant Automorphisms can be used is presented. Two of the solutions are presented in detail in section (V) alongside with tables with all the other solutions. A table concerning the existence of homothetic vector fields, and additional Killing vector fields for specific range of values of the parameters is provided. Also, a table with the invariant relations and the number of essential constants for each family, can be found. At the end of this section some remarks can be found about the solutions. Lastly, a discussion of the overall results is given.  

\section{Canonical Formalism}
In the context of canonical formalism \cite{1959PhRv..116.1322A} the line element of a ($d+1$)-dimensional manifold $M$, in a coordinate system $\Sigma$ with coordinates $(t,x^{i})$, $i=1,..,d$, acquires the form
\begin{align}\label{Le}
{ds^{2}_{(d+1)}}=\left(-N^{2}+N_{i}N^{i}\right)dt^{2}+2N_{i}dx^{i}dt+\gamma_{ij}dx^{i}dx^{j},
\end{align}
where $N(t,x^{l}), N_{i}(t,x^{l})$ are the lapse function and shift vector respectively and $\gamma_{ij}(t,x^{l})$ the metric of the $d$-dimensional sub-manifold which is given by $t=constant$.

In the light of this, the (EFE's) in vacuum are decomposed into the following equivalent set,
\begin{align}
R^{(d)}+K^{2}-K_{ij}K^{ij}=0, \hspace{0.2cm} \text{(Quadratic constraint)}\label{Qc}\\
D_{i}K-D_{j}{K_{i}}^{j}=0, \hspace{0.9cm} \text{(Linear constraint)}\label{Lc}\\
\left(\partial_{t}-{\cal{L}}_{N_{i}}\right)K_{ij}=N R^{(d)}_{ij}+N K K_{ij}-2N K_{il}{K_{j}}^{l}-D_{i}D_{j}N, \hspace{0.2cm} \text{(Dynamical equations)}\label{De}
\end{align}
where $R^{(d)}_{ij}, R^{(d)}, D_{i}$ are the Ricci tensor, the Ricci scalar and the covariant derivative constructed out of $\gamma_{ij}$ correspondingly. Also, ${\cal{L}}_{N_{i}}$ stands for the Lie derivative along the shift vector while $K_{ij}$ is the extrinsic curvature tensor given by
\begin{align}
K_{ij}=\frac{1}{2N}\left(D_{i}N_{j}+D_{j}N_{i}-\partial_{t}\gamma_{ij}\right).\nonumber
\end{align}
Note also that $K=\gamma^{ij}K_{ij}$ is the extrinsic curvature scalar.

When the manifold $M$ admits a $d$-dimensional isometry group $G$ which acts simply transitively on the $d$-dimensional sub-manifold $t=constant$, there exists an invariant basis of one-forms $\{\sigma^{\alpha}\}$ satisfying the curl relations \cite{Ryan:1975jw}
\begin{align}
d\sigma^{\alpha}=-\frac{1}{2}C^{\alpha}_{\beta\epsilon}\sigma^{\beta}\wedge\sigma^{\epsilon}\Leftrightarrow \partial_{i}\sigma^{\alpha}_{j}-\partial_{j}\sigma^{\alpha}_{i}=-C^{\alpha}_{\beta\epsilon}\sigma^{\beta}_{i}\sigma^{\epsilon}_{j},\nonumber
\end{align}
where the Greek indices run from 1 to d and $C^{\alpha}_{\beta\epsilon}$ are the structure constants of the Lie algebra of the isometry group. The sub-manifold is then called homogeneous.
Under this assumption, a coordinate system $(t,x^{i})$ exist such that the line element \eqref{Le} acquires the manifestly homogeneous form \cite{jantzen1979}
\begin{align}\label{Le1}
{ds^{2}_{(d+1)}}=\left[-N(t)^{2}+N_{\alpha}(t)N^{\alpha}(t)\right]dt^{2}+2N_{\alpha}(t)\sigma^{\alpha}_{i}(x^{l})dx^{i}dt+\gamma_{\alpha\beta}(t)\sigma^{\alpha}_{i}(x^{l})\sigma^{\beta}_{j}(x^{l})dx^{i}dx^{j},
\end{align}
and \eqref{Qc}, \eqref{Lc}, \eqref{De} reduce to ordinary differential equations, with $t$ the independent variable,
\begin{align}
R^{(d)}+K^{2}-K_{\alpha\beta}K^{\alpha\beta}=0, \hspace{0.9cm} \text{(Q. c.)}\label{Qc1}\\
{K_{\alpha}}^{\beta}C^{\alpha}_{\epsilon\beta}-{K_{\epsilon}}^{\beta}C^{\alpha}_{\beta\alpha}=0, \hspace{0.9cm} \text{(L. c.)}\label{Lc1}\\
\dot{K}_{\alpha\beta}=N R^{(d)}_{\alpha\beta}-2N K_{\alpha\epsilon}{K_{\beta}}^{\epsilon}+N K K_{\alpha\beta}-N^{\epsilon}\left(K_{\alpha\lambda}C^{\lambda}_{\epsilon\beta}+K_{\lambda\beta}C^{\lambda}_{\epsilon\alpha}\right), \hspace{0.9cm} \text{(D. e.)}\label{De1}
\end{align}
while 
\begin{align}
K_{\alpha\beta}=-\frac{1}{2N}\left(N^{\epsilon}C^{\lambda}_{\epsilon\alpha}\gamma_{\lambda\beta}+N^{\epsilon}C^{\lambda}_{\epsilon\beta}\gamma_{\lambda\alpha}+\dot{\gamma}_{\alpha\beta}\right),\nonumber
\end{align}
and
\begin{align}
R_{\alpha\beta}=-\frac{1}{2}C^{\epsilon}_{\lambda\alpha}\left(C^{\lambda}_{\epsilon\beta}+C^{\mu}_{\nu\beta}\gamma^{\lambda\nu}\gamma_{\alpha\mu}\right)+\frac{1}{4}\gamma^{\mu\nu}\gamma^{\rho\sigma}C^{\epsilon}_{\mu\rho}C^{\lambda}_{\nu\sigma}\gamma_{\alpha\lambda}\gamma_{\epsilon\beta}+\frac{1}{2}C^{\mu}_{\mu\nu}\gamma^{\nu\rho}\left(C^{\epsilon}_{\rho\alpha}\gamma_{\beta\epsilon}+C^{\epsilon}_{\rho\beta}\gamma_{\alpha\epsilon}\right).\nonumber
\end{align}

\section{Automorphism inducing Diffeomorphisms}

In the paper \cite{Christodoulakis:2000jm} a group of coordinate transformations was derived that satisfy the following conditions:\\

\begin{enumerate}
\item Preservation of sub-manifold's manifest homogeneity
\item They are symmetries of the equations \eqref{Qc1},\eqref{Lc1}, \eqref{De1} 
\end{enumerate} 
A very brief outlined of the basic idea is as follows:

For transformations of the form
\begin{align}
t\mapsto\tilde{t}=t, \hspace{2.0cm}\nonumber\\
x^{i}\mapsto{\tilde{x}}^{i}=h^{i}(t,x^{l}), \hspace{0.2cm}x^{i}=f^{i}(\tilde{t},{\tilde{x}}^{l}),\nonumber
\end{align}
the restrictions on the functions $f^{i}$, in order to satisfy the above conditions, are summarized as follows
\begin{align}
\sigma^{\alpha}_{i}\left(x^{l}\right)\frac{\partial x^{i}}{\partial {\tilde{x}}^m}=\Lambda^{\alpha}_{\beta}\left(\tilde{t}\right)\sigma^{\beta}_{m}\left({\tilde{x}}^{l}\right),\label{Lam1}\\
\sigma^{\alpha}_{i}\left(x^{l}\right)\frac{\partial x^{i}}{\partial {\tilde{t}}^m}=P^{\alpha}\left(\tilde{t}\right).\label{P1}
\end{align}
The relations \eqref{Lam1}, \eqref{P1} must be regarded as the definition of the matrix $\Lambda^{\alpha}_{\beta}$ and the vector $P^{\alpha}$ respectively.\\
The line element \eqref{Le1} can then be written as
\begin{align}
{ds^{2}_{(d+1)}}=\left[-{\tilde{N}}\left(\tilde{t}\right)^{2}+{\tilde{N}}_{\alpha}\left(\tilde{t}\right){\tilde{N}}^{\alpha}\left(\tilde{t}\right)\right]d{\tilde{t}}^{2}+2{\tilde{N}}_{\alpha}\left(\tilde{t}\right)\sigma^{\alpha}_{i}\left({\tilde{x}}^{l}\right)d{\tilde{x}}^{i}d{\tilde{t}}+{\tilde{\gamma}}_{\alpha\beta}\left(\tilde{t}\right)\sigma^{\alpha}_{i}\left({\tilde{x}}^{l}\right)\sigma^{\beta}_{j}\left({\tilde{x}}^{l}\right)d{\tilde{x}}^{i}d{\tilde{x}}^{j},\nonumber
\end{align}
with the abbreviations
\begin{equation}
\begin{aligned}\label{gammashiftlapse}
{\tilde{\gamma}}_{\alpha\beta}\left(\tilde{t}\right)={\Lambda_{\alpha}}^{\mu}\left(\tilde{t}\right){\Lambda_{\beta}}^{\nu}\left(\tilde{t}\right){\gamma}_{\mu\nu}\left(\tilde{t}\right)\\
{\tilde{N}}_{\alpha}\left(\tilde{t}\right)={\Lambda_{\alpha}}^{\mu}\left(\tilde{t}\right)\left(N_{\mu}\left(\tilde{t}\right)+P^{\nu}\left(\tilde{t}\right)\gamma_{\nu\mu}\left(\tilde{t}\right)\right)\\
\tilde{N}\left(\tilde{t}\right)=N\left(\tilde{t}\right).
\end{aligned}
\end{equation}
The existence of local solutions to the equations \eqref{Lam1},\eqref{P1} is guaranteed by the Frobenious theorem if the following necessary and sufficient conditions hold:
\begin{align}
{\Lambda^{\alpha}}_{\mu}\left(\tilde{t}\right)C^{\mu}_{\beta\nu}=C^{\alpha}_{\mu\sigma}{\Lambda^{\mu}}_{\beta}\left(\tilde{t}\right){\Lambda^{\sigma}}_{\nu}\left(\tilde{t}\right),\label{Aut}\\
{{\dot{\Lambda}}^{\alpha}}_{\!\!\!\quad\beta}\left(\tilde{t}\right)={\Lambda^{\mu}}_{\beta}\left(\tilde{t}\right)C^{\alpha}_{\mu\nu}P^{\nu}\left(\tilde{t}\right),\label{P2}
\end{align}
where the dot stands for differentiation with respect to $t$. The solutions of \eqref{Aut},\eqref{P2} form a group. 

Due to the transformation of the shift vector under the previous group, we can always choose the vector $P^{\nu}$ such that the shift vector in the transformed system is zero. Thus, both the line element and the (EFE's) acquire a simpler form.
\begin{align}
{ds^{2}_{(d+1)}}=-N\left(t\right)^{2}dt^{2}+\gamma_{\alpha\beta}(t)\sigma^{\alpha}_{i}\left(x^{l}\right)\sigma^{\beta}_{j}\left(x^{l}\right)dx^{i}dx^{j},\nonumber
\end{align}
\begin{align}
R^{(d)}+K^{2}-K_{\alpha\beta}K^{\alpha\beta}=0, \hspace{0.9cm} \text{(Q. c.)}\label{Qc2}\\
{K_{\alpha}}^{\beta}C^{\alpha}_{\epsilon\beta}-{K_{\epsilon}}^{\beta}C^{\alpha}_{\beta\alpha}=0, \hspace{0.9cm} \text{(L. c.)}\label{Lc2}\\
\dot{K}_{\alpha\beta}=N R^{(d)}_{\alpha\beta}-2N K_{\alpha\epsilon}{K_{\beta}}^{\epsilon}+N K K_{\alpha\beta}, \hspace{0.9cm} \text{(D. e.)}\label{De2}
\end{align}
while 
\begin{align}\label{Extc2}
K_{\alpha\beta}=-\frac{1}{2N}\dot{\gamma}_{\alpha\beta}.
\end{align}
The above system of equations still admits the sub-group of constant Automorphisms,
\begin{align}
{\Lambda^{\alpha}}_{\mu}C^{\mu}_{\beta\nu}=C^{\alpha}_{\mu\sigma}{\Lambda^{\mu}}_{\beta}{\Lambda^{\sigma}}_{\nu},\nonumber
\end{align}
which can also be found as "rigid" symmetries \cite{Coussaert:1993ti}.
Given the structure constants of the group, the matrix ${\Lambda^{\alpha}}_{\mu}$ is determined. The remaining non-zero elements of $\Lambda$ provide the dimension of the constant Automorphisms group.

\section{Type $4A_{1}$}

In the present work, we are interested in the case of a five-dimensional manifold with a four-dimensional homogeneous sub-manifold of Type $4A_{1}$. The structure constants for this Type are 
\begin{align}
C^{\alpha}_{\beta\mu}=0, \hspace{0.2cm} \forall \alpha,\beta,\mu=1,2,3,4.\nonumber
\end{align} 
Under the previous assumption the linear constraints \eqref{Lc2} are identically satisfied and $R^{\left(4\right)}_{\alpha\beta}=0$, $R^{\left(4\right)}=0$. Also, the Automorphisms equation is identically satisfied which implies that ${\Lambda^{\alpha}}_{\beta}$$\in$$GL(4,\mathbb{R})$.
The remaining (EFE's) \eqref{Qc2}, \eqref{De2} are
\begin{align}
K^{2}-K_{\alpha\beta}K^{\alpha\beta}=0, \hspace{0.9cm} \label{Qc3}\\
\dot{K}_{\alpha\beta}+2N K_{\alpha\epsilon}{K_{\beta}}^{\epsilon}-N K K_{\alpha\beta}=0. \label{De3}
\end{align}
By using \eqref{Extc2} and the gauge choice $N=\sqrt{\text{Det}{(\gamma)}}$ the dynamical equations \eqref{De3} are integrated 
\begin{align}
\partial_{t}\left(\gamma^{\alpha\rho}\dot{\gamma}_{\rho\beta}\right)=0\Rightarrow\hspace{0.5cm}
{\dot{\gamma}}_{\alpha\beta}={\theta^{\mu}}_{\alpha}\gamma_{\beta\mu}\hspace{0.5cm} \text{or in matrix form}\hspace{0.5cm} \dot{\gamma}=\theta^{T}\gamma, \label{geq}
\end{align}
where ${\theta^{\mu}}_{\alpha}$ is some constant matrix.\\
On the other hand, the quadratic constraint \eqref{Qc3} becomes a relation for the $\theta$ matrix.
\begin{align}
Tr(\theta^{2})-(Tr(\theta))^2=0.\nonumber
\end{align} 
The solution of \eqref{geq} is
\begin{align}
\gamma=e^{t\theta^{T}}c,\nonumber
\end{align}
where $c$ is some real, constant matrix corresponding to the value of $\gamma$ at $t=0$.\\
The matrix $\theta$ has $16$ constant elements, thus calculating the exponential is quite difficult.
This is the point where we can use the rigid symmetries in order to simplify it.\\
The action of constant Automorphisms on $\gamma$ is 
\begin{align}
\gamma=\Lambda^{T}\tilde{\gamma}\Lambda.\nonumber
\end{align}
If we use it in \eqref{geq}, the following equation holds
\begin{align}
\dot{\tilde{\gamma}}={\tilde{\theta}}^{T}\tilde{\gamma},\nonumber
\end{align}
if and only if $\theta$ transforms as follows
\begin{align}
{\tilde{\theta}}^{T}=\left({\Lambda^{T}}\right)^{-1}\theta^{T}\Lambda^{T}.\nonumber
\end{align}
Since $\Lambda\in$$GL$$\left(4,\mathbb{R}\right)$ the degree of simplification that can be achieved for the matrix $\theta$  depends upon it's eigenvalues. In the light of this, there are families of different solutions which altogether form the complete space of vacuum solutions for the Type $4A_{1}$. When there are only real eigenvalues the $\theta$ matrix transforms into it's Jordan canonical form, while when complex eigenvalues exist, acquires it's canonical rational form. A table is presented with all the possible cases and the form of the $\theta$ matrix in each one of them.

\begin{center}
\begin{table}
\begin{tabular}{ |c|c| } 
\hline
\textbf{Eigenvalues} & \textbf{Form of the matrix} \\
\hline
Four real and different eigenvalues & $
\theta=\begin{pmatrix}
p_{1} & 0 & 0 & 0 \\
0 & p_{2} & 0 & 0 \\
0 & 0 & p_{3} & 0 \\
0 & 0 & 0 & p_{4}
\end{pmatrix}
$ \\ 
\arrayrulecolor{blue}\hline \arrayrulecolor{black}
Four real eigenvalues with two of them equal & $
\theta=\begin{pmatrix}
p_{1} & 0 & 0 & 0 \\
0 & p_{2} & 0 & 0 \\
0 & 0 & p_{3} & 1 \\
0 & 0 & 0 & p_{3}
\end{pmatrix}
$ \\ 
\arrayrulecolor{blue}\hline \arrayrulecolor{black}
Four real eigenvalues with three of them equal & $
\theta=\begin{pmatrix}
p_{1} & 0 & 0 & 0 \\
0 & p_{2} & 1 & 0 \\
0 & 0 & p_{2} & 1 \\
0 & 0 & 0 & p_{2}
\end{pmatrix}
$ \\ 
\arrayrulecolor{blue}\hline \arrayrulecolor{black}
Four real and equal eigenvalues & $
\theta=\begin{pmatrix}
p_{1} & 1 & 0 & 0 \\
0 & p_{1} & 1 & 0 \\
0 & 0 & p_{1} & 1 \\
0 & 0 & 0 & p_{1}
\end{pmatrix}
$ \\ 
\arrayrulecolor{blue}\hline \arrayrulecolor{black}
Two, real, different and two complex conjugate eigenvalues & $
\theta=\begin{pmatrix}
p_{1} & 0 & 0 & 0 \\
0 & p_{2} & 0 & 0 \\
0 & 0 & p_{3} & p_{4} \\
0 & 0 & -p_{4} & p_{3}
\end{pmatrix}
$ \\ 
\arrayrulecolor{blue}\hline \arrayrulecolor{black}
Two, real, equal and two complex conjugate eigenvalues & $
\theta=\begin{pmatrix}
p_{1} & 1 & 0 & 0 \\
0 & p_{1} & 0 & 0 \\
0 & 0 & p_{2} & p_{3} \\
0 & 0 & -p_{3} & p_{2}
\end{pmatrix}
$ \\ 
\arrayrulecolor{blue}\hline \arrayrulecolor{black}
Two pairs of complex conjugate eigenvalues & $
\theta=\begin{pmatrix}
p_{1} & p_{2} & 0 & 0 \\
-p_{2} & p_{1} & 0 & 0 \\
0 & 0 & p_{3} & p_{4} \\
0 & 0 & -p_{4} & p_{3}
\end{pmatrix}
$ \\ 
\arrayrulecolor{blue}\hline \arrayrulecolor{black}
\end{tabular}
\end{table}
\end{center}

\newpage

\section{Solutions}

We present the detail calculations concerning only two of the seven different families of solutions.

\subsection{Four, real eigenvalues with two of them equal}

In this case the matrix $\theta$ is transformed into it's Jordan canonical form
\begin{align}
\theta=\begin{pmatrix}
p_{1} & 0 & 0 & 0 \\
0 & p_{2} & 0 & 0 \\
0 & 0 & p_{3} & 1 \\
0 & 0 & 0 & p_{3}
\end{pmatrix},\nonumber
\end{align}
with $p_{i}, \left(i=1,2,3\right)$ the eigenvalues. The four-dimensional line element becomes
\begin{align}
ds_{(4)}^{2}=k_{1}e^{p_{1}t}dx^{2}+k_{2}e^{p_{2}t}dy^{2}+2k_{3}e^{p_{3}t}dz\,dw+\left(k_{4}+k_{3}t\right)e^{p_{3}t}dw^{2},\nonumber
\end{align}
while the quadratic constraint reduces to
\begin{align}\label{qc2}
p_{1}p_{2}+2p_{1}p_{3}+2p_{2}p_{3}+p_{3}^{2}=0.
\end{align}
From \eqref{qc2} we observe that the restriction $\left(p_{3}\neq{0}\right)$ is required, otherwise it would lead to either $\left(p_{1}=0\right)$ or $\left(p_{2}=0\right)$, which would contradict our original statement $\left(p_{1}\neq p_{2} \neq p_{3}\right)$ implied in this section. If we divide the constraint by $p_{3}^{2}$ and form the ratios of the eigenvalues $\left(\alpha=\frac{p_{1}}{p_{3}},\beta=\frac{p_{2}}{p_{3}}\right)$ the equation becomes
\begin{align}\label{rat2}
1+2\alpha+2\beta+\alpha\beta=0.
\end{align}
For a vacuum solution we solve \eqref{rat2} with respect to one of the constants, let us choose $\alpha$.
\begin{align}
\alpha=-\frac{1+2\beta}{2+\beta}.\nonumber
\end{align}
For the branch $(\beta=-2)$, the constraint equation would be $\left(-3=0, \forall \alpha\right)$ which is invalid. Also we exclude the values $\left(\beta_{1}=-2-\sqrt{3},\beta_{2}=-2+\sqrt{3}\right)$ because they lead to $\left(\alpha=\beta\right)$. Altogether we have the following restrictions on the values of the ratio
\begin{align}
\beta\in\mathbb{R}-\left\{-2,-2+\epsilon\sqrt{3}\right\},\nonumber
\end{align}
where the symbol $\epsilon$ stands for $\epsilon=\pm1$ and will be used from now on wherever is needed.
The five-dimensional metric acquires the form
\begin{align}
ds_{(5)}^{2}=k_{1}k_{2}k_{3}^{2}e^{\phi_{3}p_{3}t}dt^{2}+k_{1}e^{\phi_{4}p_{3}t}dx^{2}+k_{2}e^{\beta p_{3}t}dy^{2}+2k_{3}e^{p_{3}t}dz\,dw+\left(k_{4}+k_{3}t\right)e^{p_{3}t}dw^{2},\nonumber
\end{align}
where $\left(\phi_{3}=\frac{3+2\beta+\beta^{2}}{2+\beta},\phi_{4}=-\frac{1+2\beta}{2+\beta}\right)$.
The eigenvalues of this metric are
\begin{align}
\lambda_{\mu}=\left(k_{1}k_{2}k_{3}^{2}e^{\phi_{3}p_{3}t},k_{1}e^{\phi_{4}p_{3}t},k_{2}e^{\beta p_{3}t},\lambda_{4},\lambda_{5}\right),\nonumber
\end{align}
with
\begin{align}
\lambda_{4}=\frac{1}{2}e^{p_{3}t}\left(k_{4}+k_{3}t-\sqrt{k_{4}^{2}+2k_{3}k_{4}t+k_{3}^{2}\left(4+t^{2}\right)}\right),\nonumber\\
\lambda_{5}=\frac{1}{2}e^{p_{3}t}\left(k_{4}+k_{3}t+\sqrt{k_{4}^{2}+2k_{3}k_{4}t+k_{3}^{2}\left(4+t^{2}\right)}\right).\nonumber
\end{align}
The signs of $(k_{3},k_{4})$ affect only the eigenvalues $(\lambda_{4},\lambda_{5})$, but always one is positive and the other is negative $\left(\forall t\in\mathbb{R}\right)$. Overall, the signature of the metric depends only on the signs of $(k_{1},k_{2})$. Only one case will be presented.

\subsubsection{$k_{i}>0, \forall \left(i=1,2\right)$}

By performing a real coordinate transformation and a redefinition of $p_{3}$,
\begin{align}
t=\frac{\tilde{t}}{p_{3}},\hspace{0.1cm} x=\frac{\sqrt{k_{2}}k_{3}}{p_{3}}\tilde{x},\hspace{0.1cm} y=\frac{\sqrt{k_{1}}k_{3}}{p_{3}}\tilde{y},\hspace{0.1cm} z=\sqrt{\frac{k_{1}k_{2}}{k_{3}}}\frac{1}{2p_{3}^{3/2}}\left(k_{4}p_{3}\tilde{w}-2k_{3}\tilde{z}\right),\hspace{0.1cm} w=-\sqrt{\frac{k_{1}k_{2}k_{3}}{p_{3}}}\tilde{w},\nonumber
\end{align}
\begin{align}
p_{3}=\sqrt{\frac{k_{1}k_{2}}{m}}k_{3},\nonumber
\end{align}
the five-dimensional line element becomes
\begin{align}
ds^{2}_{(5)}=e^{\phi_{3}t}dt^{2}+e^{\phi_{4}t}dx^{2}+e^{\beta t}dy^{2}+2e^{t}dz\,dw+t\,e^{t}dw^{2},\nonumber
\end{align}
where the constant m appearing in the redefinition of $p_{3}$, was absorbed due to the existence of the homothetic vector field
\begin{align}
\xi_{h}=\left[1,\frac{2+\beta}{2}x,\frac{3y}{4+2\beta},\frac{z\left(1+\beta+\beta^{2}\right)-w\left(2+\beta\right)}{2\left(2+\beta\right)},\frac{w\left(1+\beta+\beta^{2}\right)}{2\left(2+\beta\right)}\right].\nonumber
\end{align}

\textbf{Signature:}

In order to adjudicate for the signature of this non-diagonal metric we have to find it's eigenvalues
\begin{align}
\lambda_{\mu}=\left[e^{\phi_{3}t},e^{\phi_{4}t},e^{\beta t},\frac{1}{2}e^{t}\left(t-\sqrt{4+t^{2}}\right),\frac{1}{2}e^{t}\left(t+\sqrt{4+t^{2}}\right)\right].\nonumber
\end{align}
It is easy to see that $\left(\lambda_{4}<0, \forall t\in\mathbb{R}\right)$ while all the others are positive. The signature is Lorentzian \textit{\textbf{s}}$=(1,4)$ and the coordinate with the "time" character is either z or w.\\

\textbf{AKVF:}

This solution admits additional Killing vector fields for specific values of the ratios.

\begin{center}
    \begin{tabular}{ | c | c | c |}
    \hline
     $\alpha$ & $\beta$ & $\xi_{5}$ \\ \hline
     $1$ & $-1$ & $\left(0,w,0,-x,0\right)$ \\ \hline
     $-1$ & $1$ & $\left(0,0,w,-y,0\right)$
 \\
    \hline
    \end{tabular}
\end{center}

\subsection{Two, real, equal and two complex conjugate eigenvalues}

The matrix $\theta$ transforms into it's canonical rational form
\begin{align}
\theta=\begin{pmatrix}
p_{1} & 1 & 0 & 0 \\
0 & p_{1} & 0 & 0 \\
0 & 0 & p_{2} & p_{3} \\
0 & 0 & -p_{3} & p_{2}
\end{pmatrix},\nonumber
\end{align}
with $\left(p_{2},p_{3}\right)$ the real and imaginary parts of the complex eigenvalues correspondingly. The four-dimensional line element is
\begin{align}
ds_{(4)}^{2}=2k_{1}e^{p_{1}t}dx\,dy+\left(k_{2}+k_{1}t\right)e^{p_{1}t}dy^{2}+e^{p_{2}t}\left[k_{3}\cos(p_{3}t)+k_{4}\sin(p_{3}t)\right]\left(-dz^{2}+dw^{2}\right)\nonumber\\+2e^{p_{2}t}\left[k_{4}\cos(p_{3}t)-k_{3}\sin(p_{3}t)\right]dz\,dw,\nonumber
\end{align}
and the quadratic constraint reads
\begin{align}\label{qc6}
p_{1}^{2}+4p_{1}p_{2}+p_{2}^{2}+p_{3}^{2}=0.
\end{align}
The imaginary part of the complex eigenvalue has to be different from zero, $p_{3}\neq{0}$. If we divide \eqref{qc6} with $p_{3}^{2}$ the ratios are formed, $\left(\alpha=\frac{p_{1}}{p_{3}},\beta=\frac{p_{2}}{p_{3}}\right)$.
\begin{align}\label{rat5}
1+\alpha^{2}+4\alpha\beta+\beta^{2}=0.
\end{align}
For vacuum solutions we choose to solve \eqref{rat5} with respect to $\alpha$.
\begin{align}
\alpha=-2\beta+\epsilon\sqrt{-1+3\beta^{2}},\hspace{0.2cm} \beta\in\mathbb{R}-\left(-\frac{1}{\sqrt{3}},\frac{1}{\sqrt{3}}\right).\nonumber
\end{align}
This restriction on the ratio $\beta$ was imposed because $\alpha$ has to be real. Also, the values for which $\left(\alpha=\beta\right)$ are rejected.
\begin{align}
-2\beta+\epsilon\sqrt{-1+3\beta^{2}}=\beta\Leftrightarrow\nonumber\\
\beta=i\frac{\epsilon}{\sqrt{6}}.\nonumber
\end{align}
This relation is already satisfied.

The five-dimensional line element is 
\begin{align}
ds_{(5)}^{2}=-k_{1}^{2}\left(k_{3}^{2}+k_{4}^{2}\right)e^{\phi_{7}p_{3}t}dt^{2}+2k_{1}e^{\phi_{8}p_{3}t}dx\,dy+\left(k_{2}+k_{1}t\right)e^{\phi_{8}p_{3}t}dy^{2}\nonumber\\
+e^{\beta p_{3}t}\left[k_{3}\cos\left(p_{3}t\right)+k_{4}\sin\left(p_{3}t\right)\right]\left(-dz^{2}+dw^{2}\right)
+2e^{\beta p_{3}t}\left[k_{4}\cos\left(p_{3}t\right)-k_{3}\sin\left(p_{3}t\right)\right]dz\,dw,\nonumber
\end{align}
where $\left(\phi_{7}=-2\beta+2\epsilon\sqrt{-1+3\beta^{2}}, \phi_{8}=-2\beta+\epsilon\sqrt{-1+3\beta^{2}}\right)$ and it's eigenvalues
\begin{align}
\lambda_{\mu}=\left[-k_{1}^{2}\left(k_{3}^{2}+k_{4}^{2}\right)e^{\phi_{7}p_{3}t},\lambda_{2},\lambda_{3},-\sqrt{k_{3}^{2}+k_{4}^{2}}e^{\beta p_{3}t},\sqrt{k_{3}^{2}+k_{4}^{2}}e^{\beta p_{3}t}\right],\nonumber
\end{align}
with 
\begin{align}
\lambda_{2}=\frac{1}{2}e^{\phi_{8}p_{3}t}\left(k_{2}+k_{1}t-\sqrt{k_{2}^{2}+2k_{1}k_{2}t+k_{1}^{2}\left(4+t^{2}\right)}\right),\nonumber\\
\lambda_{3}=\frac{1}{2}e^{\phi_{8}p_{3}t}\left(k_{2}+k_{1}t+\sqrt{k_{2}^{2}+2k_{1}k_{2}t+k_{1}^{2}\left(4+t^{2}\right)}\right).\nonumber
\end{align}
The eigenvalues $(\lambda_{2},\lambda_{3})$ depend on the signs of $(k_{1},k_{2})$, but always one is positive and the other is negative, so they don't affect the signature of the metric. Also, it is easy to observe that neither $(k_{3},k_{4})$ affect the signature. Therefore, there is only one case. With the coordinate transformation and the redefinition of $p_{3}$
\begin{align}
t=\frac{\tilde{t}}{p_{3}},\hspace{0.1cm} x=-\sqrt{\frac{k_{3}^{2}+k_{4}^{2}}{k_{1}}}\frac{1}{2p_{3}^{3/2}}\left(2k_{1}\tilde{x}+k_{2}p_{3}\tilde{y}\right), \hspace{0.1cm} y=\sqrt{\frac{k_{1}\left(k_{3}^{2}+k_{4}^{2}\right)}{p_{3}}}\tilde{y},\nonumber\\
z=\frac{k_{1}\left[-k_{4}\tilde{w}+\left(-k_{3}+\sqrt{k_{3}^{2}+k_{4}^{2}}\right)\tilde{z}\right]}{\sqrt{2\left(-k_{3}+\sqrt{k_{3}^{2}+k_{4}^{2}}\right)}p_{3}}, \hspace{0.1cm} w=\frac{k_{1}\left[k_{4}\tilde{z}+\left(-k_{3}+\sqrt{k_{3}^{2}+k_{4}^{2}}\right)\tilde{w}\right]}{\sqrt{2\left(-k_{3}+\sqrt{k_{3}^{2}+k_{4}^{2}}\right)}p_{3}},\nonumber\\
p_{3}=\sqrt{k_{1}^{2}\frac{k_{3}^{2}+k_{4}^{2}}{m}},\nonumber
\end{align}
the line element simplifies to 
\begin{align}
ds_{(5)}^{2}=-e^{\phi_{7}t}dt^{2}-2e^{\phi_{8}t}dx\,dy+te^{\phi_{8}t}dy^{2}+\cos{t}\,e^{\beta t}\left(dz^{2}-dw^{2}\right)+2\sin{t}\,e^{\beta t}dz\,dw,\nonumber
\end{align}
with homothetic vector field
\begin{align}
\xi_{h}=\left(1,\frac{y+\epsilon x\sqrt{-1+3\beta^{2}}}{2},\frac{\epsilon y\sqrt{-1+3\beta^{2}}}{2},\xi_{h4},\xi_{h5}\right),\nonumber\\
\xi_{h4}=-\frac{w+3\beta z-2z\epsilon\sqrt{-1+3\beta^{2}}}{2},\nonumber\\
\xi_{h5}=\frac{z-3\beta w+2w\epsilon\sqrt{-1+3\beta^{2}}}{2}.\nonumber
\end{align}

\textbf{Signature:}

The eigenvalues are
\begin{align}
\lambda_{\mu}=\left(-e^{\phi_{7}t},\frac{t-\sqrt{4+t^{2}}}{2}e^{\phi_{8}t},\frac{t+\sqrt{4+t^{2}}}{2}e^{\phi_{8}t},e^{\beta t},-e^{\beta t}\right).\nonumber
\end{align}
The signature is \textit{\textbf{s}}$=(3,2)$ with "time" coordinates either $(t,x,w)$ or $(y,z)$.\\

\textbf{AKVF:}

For $\left(\alpha=-\frac{\epsilon}{\sqrt{2}},\beta=\frac{\epsilon}{\sqrt{2}}\right)$ the AKVF is
\begin{align}
\xi_{5}=\left(1,\frac{y+\frac{\epsilon}{\sqrt{2}}x}{2},\frac{\epsilon y}{2\sqrt{2}},-\frac{2w+\epsilon z\sqrt{2}}{4},\frac{2z-\epsilon w\sqrt{2}}{4}\right).\nonumber
\end{align}

\newpage

At this point we present the seven tables corresponding to the seven families of solutions. Also, a table in which for every solution the additional Killing vector fields (AKVF) and the homothetic field can be found. Finally, a table concerning the invariant relations and the number of essential constants for each family is given.

\subsection{Tables}

\begin{table}[htp]
    \makegapedcells
    \centering
    \resizebox{\textwidth}{!}{
\begin{tabular}{|c|c|c|}
    \hline
\multicolumn{3}{ |c| }{\textbf{Four, real and different eigenvalues}} \\
\hline
Name of solution & Line element & Signature/"Time" coordinate(s)  \\
\hline
$s_{1}$ & $ ds^{2}_{(5)}=-e^{\phi_{1}t}dt^{2}+e^{t}dx^{2}+e^{\phi_{2}t}dy^{2}+e^{\beta t}dz^{2}+e^{\gamma t}dw^{2}$ & $(1,4)\,/\,t$ \\
\hline
$s_{2}$ & $ds^{2}_{(5)}=e^{\phi_{1}t}dt^{2}+e^{t}dx^{2}+e^{\phi_{2}t}dy^{2}+e^{\beta t}dz^{2}-e^{\gamma t}dw^{2}$ & $(1,4)\,/\,w$ \\
\hline
$s_{3}$ & $ds^{2}_{(5)}=-e^{\phi_{1}t}dt^{2}+e^{t}dx^{2}+e^{\phi_{2}t}dy^{2}-e^{\beta t}dz^{2}-e^{\gamma t}dw^{2}$ & $(3,2)$\,/\,$\left(t,z,w\right)$ or $\left(x,y\right)$\\
\hline
$s_{4}$ & $ds^{2}_{(5)}=e^{\phi_{1}t}dt^{2}+e^{t}dx^{2}-e^{\phi_{2}t}dy^{2}-e^{\beta t}dz^{2}-e^{\gamma t}dw^{2}$ & $(3,2)$\,/\,$\left(y,z,w\right)$ or $\left(t,x\right)$\\
\hline
$s_{5}$ & $ds^{2}_{(5)}=-\left(e^{\phi_{1}t}dt^{2}+e^{t}dx^{2}+e^{\phi_{2}t}dy^{2}+e^{\beta t}dz^{2}+e^{\gamma t}dw^{2}\right)$ & $(0,5)$\,/\,none\\
\hline
\end{tabular}
}
\caption{The following abbreviations were used $\left(\phi_{1}={\frac{1+\beta+\gamma+\beta\gamma+\beta^{2}+\gamma^{2}}{1+\beta+\gamma}}, \phi_{2}=-\frac{\beta+\gamma+\beta\gamma}{1+\beta+\gamma}\right)$. The restrictions on the ratios of the eigenvalues are $\left(\beta\neq\left\{-1-\gamma,-1-\gamma+\epsilon\sqrt{\gamma ^2+\gamma +1},-\frac{\gamma\left(2+\gamma\right)}{1+2\gamma}\right\},\forall \gamma \in \mathbb{R}-\left\{-\frac{1}{2}\right\}\right).$}
\end{table}

\begin{table}[htp]
    \makegapedcells
    \centering
    \resizebox{\textwidth}{!}{
\begin{tabular}{|c|c|c|}
    \hline
\multicolumn{3}{ |c| }{\textbf{Four, real eigenvalues with two of them equal}} \\
\hline
Name of solution & Line element & Signature/"Time" coordinate(s)  \\
\hline
$s_{6}$ & $ ds^{2}_{(5)}=e^{\phi_{3}t}dt^{2}+e^{\phi_{4}t}dx^{2}+e^{\beta t}dy^{2}+2e^{t}dz\,dw+t\,e^{t}dw^{2}$ & $(1,4)$\,/\,$z$ or $w$ \\
\hline
$s_{7}$ & $ds^{2}_{(5)}=-e^{\phi_{3}t}dt^{2}+e^{\phi_{4}t}dx^{2}-e^{\beta t}dy^{2}+2e^{t}dz\,dw+t\,e^{t}dw^{2}$ & $(3,2)$\,/\,$\left(t,y,z\right)$or$\left(x,z\right)$ \\
\hline
$s_{8}$ & $ds^{2}_{(5)}=e^{\phi_{3}t}dt^{2}-e^{\phi_{3}t}dx^{2}-e^{\beta t}dy^{2}+2e^{t}dz\,dw+t\,e^{t}dw^{2}$ & $(3,2)$\,/\,$\left(x,y,z\right)$or$\left(t,z\right)$\\
\hline
\end{tabular}
    }
\caption{The following abbreviations were used $\left(\phi_{3}=\frac{3+2\beta+\beta^{2}}{2+\beta},\phi_{4}=-\frac{1+2\beta}{2+\beta}\right)$. The restrictions on the ratio of the eigenvalues are $\beta \in \mathbb{R}-\left\{-2,-2+\epsilon\sqrt{3}\right\}$.}
\end{table}

\begin{table}[htp]
    \makegapedcells
    \centering
    \resizebox{\textwidth}{!}{
\begin{tabular}{|c|c|c|}
    \hline
\multicolumn{3}{ |c| }{\textbf{Four, real eigenvalues with three of them equal}} \\
\hline
Name of solution & Line element & Signature/"Time" coordinate(s)  \\
\hline
$s_{9}$ & $ ds^{2}_{(5)}=e^{\left(1+3\alpha\right)t}dt^{2}+e^{t}dx^{2}+2e^{\alpha t}dy\,dw+e^{\alpha t}dz^{2}+2t\,e^{\alpha t}dz\,dw+\frac{1}{2}t^{2}e^{\alpha t}dw^{2}$ & $(1,4)\,/\,y$ \\
\hline
$s_{10}$ & $ds^{2}_{(5)}=-e^{\left(1+3\alpha\right)t}dt^{2}+e^{t}dx^{2}-2e^{\alpha t}dy\,dw-e^{\alpha t}dz^{2}+2t\,e^{\alpha t}dz\,dw-\frac{1}{2}t^{2}e^{\alpha t}dw^{2}$ & $(3,2)$\,/\,$\left(t,z,w\right)$or$\left(x,y\right)$ \\
\hline
$s_{11}$ & $ds^{2}_{(5)}=e^{\left(1+3\alpha\right)t}dt^{2}-e^{t}dx^{2}-2e^{\alpha t}dy\,dw-e^{\alpha t}dz^{2}+2t\,e^{\alpha t}dz\,dw-\frac{1}{2}t^{2}e^{\alpha t}dw^{2}$ & $(3,2)$\,/\,$\left(x,z,w\right)$ or $\left(t,y\right)$\\
\hline
\end{tabular}
    }
\caption{There are only two possible values for the ratio of the eigenvalues for which we acquire a vacuum solution, $\left(\alpha=0\,\text{or}\,a=-1\right)$.}
\end{table}

\begin{table}[htp]
    \makegapedcells
    \centering
    \resizebox{\textwidth}{!}{
\begin{tabular}{|c|c|c|}
    \hline
\multicolumn{3}{ |c| }{\textbf{Four, real and equal eigenvalues}} \\
\hline
Name of solution & Line element & Signature/"Time" coordinate(s)  \\
\hline
$s_{12}$ & $ ds_{(5)}^{2}=-dt^{2}-2dx\,dw-2dy\,dz-2t\,dy\,dw+t\,dz^{2}+t^{2}dz\,dw+\frac{t^{3}}{6}dw^{2}$ & $(3,2)$\,/\,$\left(t,z,w\right)$or$\left(x,y\right)$ \\
\hline
$s_{13}$ & $ds_{(5)}^{2}=-\left(dt^{2}+2dx\,dw+2dy\,dz+2t\,dy\,dw+t\,dz^{2}+t^{2}dz\,dw+\frac{t^{3}}{6}dw^{2}\right)$ & $(3,2)$\,/\,$\left(x,z,w\right)$or$\left(t,y\right)$ \\
\hline
\end{tabular}
    }
    \caption{\qquad\qquad\qquad\qquad\qquad\qquad\qquad\qquad\qquad\qquad\qquad\qquad\qquad\qquad\qquad\qquad\qquad\qquad\qquad\qquad\qquad}
\end{table}

\begin{table}[htp]
    \makegapedcells
    \centering
    \resizebox{\textwidth}{!}{
\begin{tabular}{|c|c|c|}
    \hline
\multicolumn{3}{ |c| }{\textbf{Two, real, different and two complex conjugate eigenvalues}} \\
\hline
Name of solution & Line element & Signature/"Time" coordinate(s)  \\
\hline
$s_{14}$ & $ds_{(5)}^{2}=e^{\phi_{5}t}dt^{2}+e^{\phi_{6}t}dx^{2}+e^{\beta t}dy^{2}+\cos{t}\,e^{\gamma t}\left(dz^{2}-dw^{2}\right)+2\sin{t}\,e^{\gamma t}dz\,dw$ & $(1,4)\,/\,w$ \\
\hline
$s_{15}$ & $ds_{(5)}^{2}=-e^{\phi_{5}t}dt^{2}+e^{\phi_{6}t}dx^{2}-e^{\beta t}dy^{2}+\cos{t}\,e^{\gamma t}\left(dz^{2}-dw^{2}\right)+2\sin{t}\,e^{\gamma t}dz\,dw$ & $(3,2)$\,/\,$\left(t,y,w\right)$or$\left(x,z\right)$ \\
\hline
$s_{16}$ & $ds_{(5)}^{2}=e^{\phi_{5}t}dt^{2}-e^{\phi_{6}t}dx^{2}-e^{\beta t}dy^{2}+\cos{t}\,e^{\gamma t}\left(dz^{2}-dw^{2}\right)+2\sin{t}\,e^{\gamma t}dz\,dw$ & $(3,2)$\,/\,$\left(x,y,w\right)$ or $\left(t,z\right)$\\
\hline
$s_{17}$ & $ds_{(5)}^{2}=e^{\alpha t}\left(dt^{2}+dx^{2}\right)+e^{-\frac{2\epsilon}{\sqrt{3}} t}dy^{2}+\cos{t}\,e^{\frac{\epsilon}{\sqrt{3}} t}\left(dz^{2}-dw^{2}\right)+2\sin{t}\,e^{\frac{\epsilon}{\sqrt{3}} t}dz\,dw$ & $(1,4)$\,/\,$w$\\
\hline
$s_{18}$ & $ds_{(5)}^{2}=e^{\alpha t}\left(-dt^{2}+dx^{2}\right)-e^{-\frac{2\epsilon}{\sqrt{3}} t}dy^{2}+\cos{t}\,e^{\frac{\epsilon}{\sqrt{3}} t}\left(dz^{2}-dw^{2}\right)+2\sin{t}\,e^{\frac{\epsilon}{\sqrt{3}} t}dz\,dw$ & $(3,2)$\,/\,$\left(t,y,w\right)$or$\left(x,z\right)$\\
\hline
$s_{19}$ & $ds_{(5)}^{2}=e^{\alpha t}\left(dt^{2}-dx^{2}\right)-e^{-\frac{2\epsilon}{\sqrt{3}} t}dy^{2}+\cos{t}\,e^{\frac{\epsilon}{\sqrt{3}} t}\left(dz^{2}-dw^{2}\right)+2\sin{t}\,e^{\frac{\epsilon}{\sqrt{3}} t}dz\,dw$ & $(3,2)$\,/\,$\left(x,y,w\right)$or$\left(t,z\right)$\\
\hline
\end{tabular}
    }
\caption{The following abbreviations were used $\left(\phi_{5}=\frac{-1+\beta^{2}+2\beta\gamma+3\gamma^{2}}{\beta+2\gamma},\phi_{6}=-\frac{1+2\beta\gamma+\gamma^{2}}{\beta+2\gamma}\right)$. The restrictions on the ratios of the eigenvalues are $\beta$$\in$$\mathbb{R}-\left\{-2\gamma,-2\gamma+\epsilon\sqrt{-1+3\gamma^{2}},-\frac{1+3\gamma^{2}}{3\gamma}\right\}$,$\forall \gamma\in \mathbb{R}-\left(-\frac{1}{\sqrt{3}},\frac{1}{\sqrt{3}}\right)$,\quad$\alpha\in \mathbb{R}$.}
\end{table}

\begin{table}[htp]
    \makegapedcells
    \centering
    \resizebox{\textwidth}{!}{
\begin{tabular}{|c|c|c|}
    \hline
\multicolumn{3}{ |c| }{\textbf{Two, real, equal and two complex conjugate eigenvalues}} \\
\hline
Name of solution & Line element & Signature/"Time" coordinate(s)  \\
\hline
$s_{20}$ & $ds_{(5)}^{2}=-e^{\phi_{7}t}dt^{2}-2e^{\phi_{8}t}dx\,dy+te^{\phi_{8}t}dy^{2}+\cos{t}\,e^{\beta t}\left(dz^{2}-dw^{2}\right)+2\sin{t}\,e^{\beta t}dz\,dw$ & $(3,2)$\,/\,$\left(t,x,w\right)$or$\left(y,z\right)$ \\
\hline
\end{tabular}
    }
\caption{The following abbreviations were used $\left(\phi_{7}=-2\beta+2\epsilon\sqrt{-1+3\beta^{2}}, \phi_{8}=-2\beta+\epsilon\sqrt{-1+3\beta^{2}}\right)$. The restrictions on the ratios of the eigenvalues are $ \beta\in\mathbb{R}-\left(-\frac{1}{\sqrt{3}},\frac{1}{\sqrt{3}}\right)$.}
\end{table}

\begin{table}[htp]
    \makegapedcells
    \centering
    \resizebox{\textwidth}{!}{
\begin{tabular}{|c|c|c|}
    \hline
\multicolumn{3}{ |c| }{\textbf{Two pairs of complex conjugate eigenvalues}} \\
\hline
Name of solution & Line element & Signature/"Time" coordinate(s)  \\
\hline
$s_{21}$ & $ds_{(5)}^{2}=-e^{\phi_{9}t}dt^{2}
+\cos\left(\beta t\right)\,e^{\phi_{10}t}\left(dx^{2}-dy^{2}\right)+2\sin\left(\beta t\right)\,e^{\phi_{10}t}dx\,dy+2\sin{t}\,e^{\gamma t}dz\,dw
+\cos{t}\,e^{\gamma t}\left(dz^{2}-dw^{2}\right)$ & $(3,2)$\,/\,$\left(t,y,w\right)$or$\left(x,z\right)$ \\
\hline
\end{tabular}
    }
\caption{The following abbreviations were used $\left(\phi_{9}=-2\gamma+2\epsilon\sqrt{-1-\beta^{2}+3\gamma^{2}}, \phi_{10}=-2\gamma+\epsilon\sqrt{-1-\beta^{2}+3\gamma^{2}}\right)$. The restrictions on the ratios of the eigenvalues are $\beta\in\left(-\sqrt{-1+3\gamma^{2}},\sqrt{-1+3\gamma^{2}}\right)$$-$$\left\{\frac{1}{2}\left(-2\gamma+\epsilon\sqrt{-2+2\gamma^{2}}\right)\right\}$,\quad$\gamma\in\mathbb{R}-\left(-\frac{1}{\sqrt{3}},\frac{1}{\sqrt{3}}\right)$.}
\end{table}

\newpage

\begin{center}
\begin{tabular}{ |c|c|c| } 
\hline
\textbf{S} & \textbf{AKVF/Values for which these appear} & \textbf{Homothetic field} \\
\hline
\multirow{3}{*}{$s_{1}$} &  $\left(0,-y,x,0,0\right)$/$\alpha=1$,\,$\beta=-\frac{1+2\gamma}{2+\gamma}$,\,$\gamma=\mathbb{R}-\{-2,-1,1\}$ & \\ \cline{2-2} 
& $\left(0,-z,0,x,0\right)$/$\alpha=-\frac{1+2\gamma}{2+\gamma}$,\,$\beta=1$,\,$\gamma=\mathbb{R}-\{-2,-1,1\}$ & \\ \cline{2-2}
&  $\left(0,-w,0,0,x\right)$/$\alpha=-\frac{1+2\beta}{2+\beta}$,\,$\beta=\mathbb{R}-\{-2,-1,1\}$,\,$\gamma=1$ & \\ 
\arrayrulecolor{red}\cline{1-2} \arrayrulecolor{black}
\multirow{3}{*}{$s_{2}$} & same as $s_{1}$ & \\ \cline{2-2} 
& same as $s_{1}$ & \\ \cline{2-2}
&  $\left(0,w,0,0,x\right)$/$\alpha=-\frac{1+2\beta}{2+\beta}$,\,$\beta=\mathbb{R}-\{-2,-1,1\}$,\,$\gamma=1$ & \\ 
\arrayrulecolor{red}\cline{1-2} \arrayrulecolor{black}
\multirow{3}{*}{$s_{3}$} &  same as $s_{1}$ & \\ \cline{2-2} 
& $\left(0,z,0,x,0\right)$/$\alpha=-\frac{1+2\gamma}{2+\gamma}$,\,$\beta=1$,\,$\gamma=\mathbb{R}-\{-2,-1,1\}$ & $\left[1,\frac{\beta^{2}+\gamma^{2}+\beta\gamma}{2(1+\beta+\gamma)}x,\frac{y(1+\beta+\gamma)}{2},\frac{z(1+\gamma+\gamma^{2})}{2(1+\beta+\gamma)},\frac{w(1+\beta+\beta^{2})}{2(1+\beta+\gamma)}\right]$ \\ \cline{2-2}
&  same as $s_{2}$ &  \\ 
\arrayrulecolor{red}\cline{1-2} \arrayrulecolor{black}
\multirow{3}{*}{$s_{4}$} &  $\left(0,y,x,0,0\right)$/$\alpha=1$,\,$\beta=-\frac{1+2\gamma}{2+\gamma}$,\,$\gamma=\mathbb{R}-\{-2,-1,1\}$ & \\ \cline{2-2} 
& same as $s_{3}$ &\\ \cline{2-2}
&  same as $s_{2}$ & \\ 
\arrayrulecolor{red}\cline{1-2} \arrayrulecolor{black}
\multirow{3}{*}{$s_{5}$} &  same as $s_{1}$ & \\ \cline{2-2} 
& same as $s_{1}$ & \\ \cline{2-2}
& same as $s_{1}$ & \\ 
\arrayrulecolor{blue}\hline \arrayrulecolor{black}
\multirow{2}{*}{$s_{6}$} &  $\left(0,w,0,-x,0\right)$/$\alpha=1$,\,$\beta=-1$ & \\ \cline{2-2} 
 & $\left(0,0,w,-y,0\right)$/$\alpha=-1$,\,$\beta=1$ & \\ 
\arrayrulecolor{red}\cline{1-2} \arrayrulecolor{black}
\multirow{2}{*}{$s_{7}$} &  same as $s_{6}$ & \multirow{2}{*}{$\left[1,\frac{2+\beta}{2}x,\frac{3y}{4+2\beta},\frac{z\left(1+\beta+\beta^{2}\right)-w\left(2+\beta\right)}{2\left(2+\beta\right)},\frac{w\left(1+\beta+\beta^{2}\right)}{2\left(2+\beta\right)}\right]$}\\ \cline{2-2} 
 & $\left(0,0,w,y,0\right)$/$\alpha=-1$,\,$\beta=1$ &\\ 
\arrayrulecolor{red}\cline{1-2} \arrayrulecolor{black}
\multirow{2}{*}{$s_{8}$} &  $\left(0,w,0,x,0\right)$/$\alpha=1$,\,$\beta=-1$ & \\ \cline{2-2} 
 & same as $s_{7}$ & \\   
\arrayrulecolor{blue}\hline \arrayrulecolor{black}
$s_{9}$ & none additional Killing vector field & \\ 
\arrayrulecolor{red}\cline{1-2} \arrayrulecolor{black}
$s_{10}$ &  none additional Killing vector field & $\left[1,\frac{3\alpha}{2}x,\frac{\left(y-z+2\alpha y\right)}{2},\frac{\left(-w+z+2\alpha z\right)}{2},\frac{1+2\alpha}{2}w\right]$\\  
\arrayrulecolor{red}\cline{1-2} \arrayrulecolor{black}
$s_{11}$ &  none additional Killing vector field & \\    
\arrayrulecolor{blue}\hline \arrayrulecolor{black}
$s_{12}$ & $\left(1,-\frac{y}{2},\frac{z}{2},-\frac{w}{2},0\right)$ & \multirow{2}{*}{$\left(t,\frac{5x}{2},\frac{3y}{2},\frac{z}{2},-\frac{w}{2}\right)$}\\ 
\arrayrulecolor{red}\cline{1-2} \arrayrulecolor{black}
$s_{13}$ &  $\left(1,-\frac{y}{2},-\frac{z}{2},-\frac{w}{2},0\right)$ &\\      
\arrayrulecolor{blue}\hline \arrayrulecolor{black}
$s_{14}$ &  & \\ 
\arrayrulecolor{red}\cline{1-1} \arrayrulecolor{black}
\multirow{4}{*}{$s_{15}$} & $\left[1,\frac{x\left(\gamma+\epsilon\sqrt{1-2\gamma^{2}}\right)}{2},\frac{y\left(\gamma-\epsilon\sqrt{1-2\gamma^{2}}\right)}{2},-\frac{w+z\gamma}{2},\frac{z-w\gamma}{2}\right]$/ & $\left[1,\frac{\beta+2\gamma}{2}x,\frac{-1+3\gamma^{2}}{2\left(\beta+2\gamma\right)}y,\xi_{h4},\xi_{h5}\right]$\\  
& $\alpha=\frac{-1+\gamma^{2}-2\epsilon\gamma\sqrt{1-2\gamma^{2}}}{\gamma+\epsilon\sqrt{1-2\gamma^{2}}}$ & $\xi_{h4}=\frac{-w\left(\beta+2\gamma\right)+z\left(-1+\beta^{2}+\beta\gamma+\gamma^{2}\right)}{2\left(\beta+2\gamma\right)}$\\ 
&  $\beta=-\gamma+\epsilon\sqrt{1-2\gamma^{2}}$ & $\xi_{h5}=\frac{z\left(\beta+2\gamma\right)+w\left(-1+\beta^{2}+\beta\gamma+\gamma^{2}\right)}{2\left(\beta+2\gamma\right)}$\\
& $-\frac{1}{\sqrt{2}}<\gamma<\frac{1}{\sqrt{2}}$ &\\
\arrayrulecolor{red}\cline{1-1} \arrayrulecolor{black}
$s_{16}$ & &\\
\hline
\end{tabular}
\end{center}

\begin{center}
\begin{tabular}{ |c|c|c| } 
\hline
\textbf{S} & \textbf{AKVF/Values for which these appear} & \textbf{Homothetic field} \\
\hline
$s_{17}$ &  & \multirow{3}{*}{$\left(1,0,y\frac{\sqrt{3}\alpha+2\epsilon}{2\sqrt{3}},\frac{3\alpha z-3w-\epsilon z\sqrt{3}}{6},\frac{3\alpha w+3z-\epsilon w\sqrt{3}}{6}\right)$}\\  
\arrayrulecolor{red}\cline{1-1} \arrayrulecolor{black}
$s_{18}$ &  $\left(1,0,\frac{\epsilon y}{\sqrt{3}},-\frac{3w+\epsilon z\sqrt{3}}{6},\frac{3z-\epsilon w\sqrt{3}}{6}\right)$/$\alpha=0$ &\\  
\arrayrulecolor{red}\cline{1-1} \arrayrulecolor{black}
$s_{19}$ &  &\\   
\arrayrulecolor{blue}\hline \arrayrulecolor{black}
\multirow{2}{*}{$s_{20}$} & $\left(1,\frac{\sqrt{2}y+\epsilon x}{2\sqrt{2}},\frac{\epsilon y}{2\sqrt{2}},-\frac{2w+\epsilon z\sqrt{2}}{4},\frac{2z-\epsilon w\sqrt{2}}{4}\right)$/ & $\left(1,\frac{y+\epsilon x\sqrt{-1+3\beta^{2}}}{2},\frac{\epsilon y\sqrt{-1+3\beta^{2}}}{2},\xi_{h4},\xi_{h5}\right)$\\
& $\left(\alpha=-\frac{\epsilon}{\sqrt{2}},\beta=\frac{\epsilon}{\sqrt{2}}\right)$ & $\xi_{h4}=-\frac{w+3\beta z-2z\epsilon\sqrt{-1+3\beta^{2}}}{2}$, $\xi_{h5}=\frac{z-3\beta w+2w\epsilon\sqrt{-1+3\beta^{2}}}{2}$\\      
\arrayrulecolor{blue}\hline \arrayrulecolor{black}
\multirow{4}{*}{$s_{21}$} & $\left(1,\frac{\lvert{\gamma}\rvert x-y\sqrt{-1+2\gamma^{2}}}{2},\frac{\lvert{\gamma}\rvert y+x\sqrt{-1+2\gamma^{2}}}{2},\xi_{4},\xi_{5}\right)$ & $\left(1,\frac{-\beta y+\epsilon x\sqrt{-1-\beta^{2}+3\gamma^{2}}}{2},\frac{\beta x+\epsilon y\sqrt{-1-\beta^{2}+3\gamma^{2}}}{2},\xi_{h4},\xi_{h5}\right)$\\
& $\xi_{4}=-\frac{w}{2}+z\left(-\frac{3\gamma}{2}+\lvert{\gamma}\rvert\right)$,\,$\xi_{5}=\frac{z}{2}+w\left(-\frac{3\gamma}{2}+\lvert{\gamma}\rvert\right)$/ & $\xi_{h4}=-\frac{w}{2}-\frac{3\gamma z}{2}+\epsilon z\sqrt{-1-\beta^{2}+3\gamma^{2}}$\\ 
& $\alpha=-2\gamma+\epsilon\lvert{\gamma}\rvert$,\,$\beta=\epsilon\sqrt{-1+2\gamma^{2}}$ & $\xi_{h5}=\frac{z}{2}-\frac{3\gamma w}{2}+\epsilon w\sqrt{-1-\beta^{2}+3\gamma^{2}}$\\
& $\gamma\in\mathbb{R}-\left(-\frac{1}{\sqrt{3}},\frac{1}{\sqrt{3}}\right)$ &\\ 
\hline
\end{tabular}
\end{center}

\newpage

\begin{center}
\begin{table}
\begin{tabular}{ |c|c|c| } 
\hline
\textbf{S} & \textbf{Invariant relations} & \textbf{Number of essential constants}\\
\hline
$s_{1}$-$s_{5}$ & $\frac{Q}{K^{5/2}}=f_{1}\left(\beta,\gamma\right)$,\hspace{0.2cm} $\frac{W}{K^{3/2}}=f_{2}\left(\beta,\gamma\right)$ & 2\\ 
\arrayrulecolor{blue}\hline \arrayrulecolor{black}
$s_{6}$-$s_{8}$ & $\frac{Q}{K^{5/2}}=f_{3}\left(\beta\right)$ & 1\\
\arrayrulecolor{blue}\hline \arrayrulecolor{black}
$s_{9}$-$s_{11}$/$\alpha=-1$ &  $\frac{Q}{K^{5/2}}=\frac{16\sqrt{2}}{3}$ & 0\\
\arrayrulecolor{blue}\hline \arrayrulecolor{black}
$s_{9}$-$s_{11}$/$\alpha=0$ &  None & 0\\
\arrayrulecolor{blue}\hline \arrayrulecolor{black}
$s_{12}$-$s_{13}$ &  None & 0\\
\arrayrulecolor{blue}\hline \arrayrulecolor{black}
$s_{14}$-$s_{16}$ & $\frac{Q}{K^{5/2}}=f_{4}\left(\beta,\gamma\right)$,\hspace{0.2cm} $\frac{W}{K^{3/2}}=f_{5}\left(\beta,\gamma\right)$ & 2\\
\arrayrulecolor{blue}\hline \arrayrulecolor{black}
$s_{17}$-$s_{19}$ & $\frac{Q}{K^{5/2}}=2\times3^{1/4}\alpha^{3/2}$ & 1\\
\arrayrulecolor{blue}\hline \arrayrulecolor{black}
$s_{20}$ & $\frac{Q}{K^{5/2}}=f_{6}\left(\beta\right)$ & 1\\
\arrayrulecolor{blue}\hline \arrayrulecolor{black}
$s_{21}$ & $\frac{Q}{K^{5/2}}=f_{7}\left(\beta,\gamma\right)$,\hspace{0.2cm} $\frac{W}{K^{3/2}}=f_{8}\left(\beta,\gamma\right)$ & 2\\
\hline
\end{tabular}
\caption{We used the abbreviations $K=R^{\mu\nu\sigma\rho}R_{\mu\nu\sigma\rho}$, $Q=g^{\mu\nu}\nabla_{\mu}K\nabla_{\nu}K$, $W=\nabla^{\lambda}R^{\mu\nu\sigma\rho}\nabla_{\lambda}R_{\mu\nu\sigma\rho}$.}
\end{table}
\end{center}

\subsection{Remarks}

Some remarks concerning the solutions found are noteworthy.
\begin{enumerate}
\item To the best of our knowledge, only the first family of solutions is known. This corresponds to the five-dimensional Kasner Type \cite{10.2307/2370192}, \cite{10.2307/1989167}.
\item For the solutions $s_{1}$-$s_{5}$ there is the possibility to choose either of the ratios to be equal to zero, suppose $\gamma=0$. These represent the four-dimensional Kasner solutions \cite{10.2307/2370192}, \cite{10.2307/1989167} embedded in a five-dimensional manifold.
\item In the case of $\left(s_{9}, s_{10}, s_{11}\right)$, for one of the two possible values of the ratio $(\alpha=0)$, the Killing vector field $\left(\xi={\partial_{y}}\right)$ satisfies the following properties,
\begin{align}
\xi_{\mu}\xi^{\mu}=0,\nonumber\\
\nabla_{\mu}\xi_{\nu}=0.\nonumber
\end{align}
This is the definition of pp-waves \cite{Kundt1961} which are solutions to (EFE's), representing strong gravitational waves propagating along null trajectories, generated by some null vector field. In our case this field is $\xi=\partial_{y}$.
\item The solutions $\left(s_{12}, s_{13}\right)$ are also pp-waves, with $\left(\xi=\partial_{x}\right)$ being the null vector field.
\item When it comes to pp-wave solutions there are no invariant relations and this is due to the fact that all the curvature scalars are zero. Therefore, in these solutions the method of curvature invariants cannot be used in order to adjudicate whether or not a constant is essential. Another method can be found in \cite{doi:10.1063/1.2338760}. In our case of course there are no constants in the pp-wave line elements and so there is no need to apply this method.  
\end{enumerate}

\section{Discussion}

As we have seen, the group of transformations that preserve the sub-manifold's manifest homogeneity provides us with a way to put the shift vector equal to zero without doing it a priori. Also, we use the gauge freedom of choosing the lapse function in order to simplify further the (EFE's). The task of finding the explicit form of the general solution of the simplified (EFE's) remained difficult even in the simple case of Type $4A_{1}$. The use of the remaining symmetry, consisting of the group of constant Automorphisms, allowed us to overcome this difficulty. This was due to the fact that the solution space was broken down into different families of solutions with the aid of the constant Automorphisms. By this procedure, we were able to find out all the known solutions as well as all the possible new, to the best of our knowledge. Of course, the solutions found span the entire space-time metrics with non-zero shift and arbitrary lapse; one only has to invert the transformations \eqref{gammashiftlapse} which enable us to reduce the initial metric to the form which has been used for the subsequent finding of the solutions.

For the future, we aim to use this method in the rest of the five-dimensional manifolds which admit a homogeneous sub-manifold of dimension four, where the group of constant Automorphisms has a lower dimension. Also a matter content may be included. Finally, as expected, in every solution the $t$$=\text{constant}$ hyper-surfaces are flat. It is worth observing that this is not true in general if we choose any other of the coordinates to be constant. It would be interesting to investigate if any of the non-flat hyper-surfaces of dimension four are solutions to the four-dimensional (EFE's), probably coupled to some appropriate matter content.

\section*{Acknowledgements}
The co-author T. Pailas thanks the General Secretariat for Research and Technology (GSRT) and the Hellenic Foundation for Research and Innovation (HFRI) of the Greek Ministry of Education for supporting his PhD fellowship.

\nocite{*}

\medskip

\bibliographystyle{unsrt}

\end{document}